\documentclass[nofootinbib, reprint, amsmath,amssymb, aps,]{revtex4-2}
\usepackage{graphicx}
\usepackage{dcolumn}
\usepackage{bm}
\usepackage[colorlinks]{hyperref}

\begin{document}


\title{Magnetic stochasticity and diffusion}

\author{Amir Jafari}
 \email{ajafari4@alumni.jh.edu}%
\author{Ethan Vishniac}%
 \email{evishni1@jhu.edu}
 \author{Vignesh Vaikundaraman}%
 \email{vvaikun1@jhu.edu}

\affiliation{%
Johns Hopkins University, Baltimore, USA\\
}%

\date{\today}

\begin{abstract}
We develop a quantitative relationship between magnetic diffusion and the level of randomness, or stochasticity, of the diffusing magnetic field in a magnetized medium. A general mathematical formulation of magnetic stochasticity in turbulence has been developed in previous work in terms of the ${\cal L}_p$-norm $S_p(t)={1\over 2}|| 1-\hat{\bf B}_l.\hat{\bf B}_L||_p$, $p$th order magnetic stochasticity of the stochastic field ${\bf B}({\bf x}, t)$, based on the coarse-grained fields, ${\bf B}_l$ and ${\bf B}_L$, at different scales, $l\neq L$. For laminar flows, stochasticity level becomes the level of field self-entanglement or spatial complexity. In this paper, we establish a connection between magnetic stochasticity $S_p(t)$ and magnetic diffusion in magnetohydrodynamic (MHD) turbulence and use a homogeneous, incompressible MHD simulation to test this prediction. Our results agree with the well-known fact that magnetic diffusion in turbulent media follows the super-linear Richardson dispersion scheme. This is intimately related to stochastic magnetic reconnection in which super-linear Richardson diffusion broadens the matter outflow width and accelerates the reconnection process.

\end{abstract}

\pacs{Valid PACS appear here}
\maketitle


\section{\label{sec:level1}Introduction}

In the early 1940s, Onsager pointed out, but never published, the remarkable fact that the velocity field in a turbulent fluid becomes H{\"o}lder singular\footnote{The complex (or real) valued function $g$ in ${\cal{R}}^n$ is H{\"o}lder continuous if two non-negative and real constants $C$ and $h$ exist such that $| g(x) - g(y) | \leq C\| x - y\|^h$ for all $x, y \in Domain(g)$. If the H{\"o}lder exponent $h$ is equal to unity, then $g$ is Lipschitz continuous. Also $g$ is called H{\"o}lder singular if $h< 1$.} in the limit of vanishing viscosity; $\nu\rightarrow 0$ (\cite{EyinkS2006}; \cite{Eyink2018}; \cite{JV2019}). This approach was based on an exact mathematical analysis of the high Reynolds-number regime of incompressible hydrodynamic turbulence. Such an analysis can be called, using a slightly more modern language, a non-perturbative renormalization group analysis \cite{Eyink2018}. Both laboratory experiments and numerical simulations (see e.g., \cite{Sreenivasan1984}; \cite{Sreenivasan1998}; and also \cite{EyinkS2006}; \cite{Eyink2018} and references therein) have confirmed that the kinetic energy dissipation rate in a fluid does not vanish in the limit of vanishing viscosity $\nu$. Such a non-vanishing limit of energy dissipation requires that space-gradients of velocity diverge in the limit $\nu \rightarrow 0$, i.e., $\nabla{\bf u}\rightarrow \infty$ ( \cite{Eyink2018}; \cite{JV2019}). This blow-up of velocity gradients resembles ultra-violet singularities encountered in quantum field theory. Therefore, hydrodynamic equations will become ill-defined in this limit as they contain velocity gradients. Turbulent magnetic fields as well face the same singularity problem in the limit when viscosity and resistivity tend to zero $\nu, \eta\rightarrow 0$ simultaneously. Consequently, MHD equations become ill-defined in this limit when the flow is turbulent. All in all, this suggests that the conventional ideal hydrodynamics (HD) and ideal MHD may be applied only if the flows remain laminar and all quantities Lipschitz-continuous. 

In a magnetized fluid, the magnetic diffusivity (resistivity) and viscosity may be small but finite. In the limit of vanishing magnetic diffusivity, the magnetic field seems to be frozen into the fluid. This magnetic flux-freezing principle is widely applied as an estimate to MHD equations in the laboratory and astrophysical systems with the presumption that ideal MHD holds to a good accuracy. With turbulence, ubiquitous in astrophysical and laboratory systems (see e.g., \cite{Eyinketal.2013}; \cite{JafariandVishniac2018}; \cite{JafariVishniac2018disks}; \cite{Jafari2019} and references therein), the velocity and magnetic fields become singular in the limit $\nu, \eta\rightarrow 0$ and ideal MHD cannot be applied. For instance, magnetic (and velocity) field lines are usually assumed to be well-defined in such approaches, however, mathematically, the existence and uniqueness of integral curves (field lines) is guaranteed only for Lipschitz continuous vector fields. What does a magnetic field line mean if the field is (H{\"o}lder) singular rather than Lipschitz-continuous?  
 
It has been shown that in the limit when viscosity of a turbulent fluid tends to zero, its Lagrangian particle trajectories become stochastic (see e.g., \cite{Bernardetal.1998}; \cite{GawedzkiandVergassola2000}; \cite{VandenEijndenandVandenEijnden2000}). Also, it turns out that magnetic field lines become stochastic in turbulent, magnetized fluids in the limit when resistivity and viscosity tend to zero simultaneously (\cite{Eyink2011}; \cite{Eyinketal.2013}; \cite{Eyink2015}; \cite{Lalescuetal.2015}). Under such circumstances, instead of the conventional magnetic flux freezing \cite{Alfven1942} a stochastic version is introduced \cite{Eyink2011}. Although the concept of a stochastic magnetic field is used in such contexts, but it is not mathematically obvious at all what a stochastic vector field really means. In other words, the notion of a stochastic variable is well-known for scalar quantities such as a fluctuating temperature or the price of certain goods in the market. However, a vector field assigns a vector, with magnitude and direction, to every point in space and time and we need a more general statistical formulation to study the randomness of a vector field and its relationship with the topology and other features of the field.

Jafari and Vishniac \cite{JV2019} presented a mathematical formulation for the stochasticity level of magnetic fields in terms of the unit vectors tangent to the renormalized fields at different coarse-graining scales. The time dependent angle between such two unit vectors at a space-time point $({\bf x}, t)$ provides a means to define a local stochasticity level; see \S \ref{s1}. The average stochasticity level in an arbitrary volume $V$ can then be defined using ${\cal L}_p$ norms. The time evolution of the stochasticity level, defined in this way, would then be associated with the topological deformations of the magnetic field. 

In the present paper, first we briefly review the concept of vector field stochasticity developed by \cite{JV2019} in \S \ref{s1}. In \S\ref{s2}, we relate magnetic diffusion to magnetic stochasticity and test the theoretically predicted relationship using the data extracted from an incompressible, homogenous MHD simulation, archived in an online, web-accessible database (\citep*{JHTDB};\citep*{JHTB1};\citep*{JHTB2}). In \S\ref{s3}, we summarize and discuss our results. In order to present a more complete discussion on magnetic diffusion, we have also added an appendix to discuss the 2-particle Richardson diffusion and the related scaling laws in MHD turbulence.

\section{Vector Field Stochasticity}\label{s1}

In order to remove the singularities of the velocity field ${\bf u}({\bf x}, t)$ or magnetic field ${\bf{B}}({\bf{x}}, t)$ in a turbulent flow, we can renormalize (coarse-grain) it at a length scale $l$ by multiplying it by a rapidly decaying function and integrating over a volume $V$. For example, for magnetic field ${\bf B}$, we have

\begin{equation}\label{1}
{\bf{B}}_l ({\bf{x}}, t)=\int_V G_l({\bf{r}})  {\bf{B}}({\bf{x+r}}, t) d^3r,
\end{equation}
where $G_l({\bf{r}}) =l^{-3} G({\bf{r}}/l) $ with $G({\bf{r}})$ being a smooth, rapidly decaying kernel, e.g., the Gaussian kernel scales as $ e^{-r^2/l^2}$. Without loss of generality, we may assume

\begin{equation}\label{2}
G({\bf{r}})\geq 0,
\end{equation}

\begin{equation}\label{3}
Lim_{|\bf r|\rightarrow \infty} G({\bf{r}})\rightarrow 0,
\end{equation}

\begin{equation}\label{4}
\int_V d^3r G({\bf{r}})=1,
\end{equation}

\begin{equation}\label{5}
\int_V d^3r \; {\bf{r}}\;G({\bf{r}})=0,
\end{equation}
and
\begin{equation}
\int_V d^3r |{\bf{r}}|^2 \;G({\bf{r}})= 1.
\end{equation}
The renormalized field ${\bf{B}}_l$ represents the average field in a parcel of fluid of length scale $l$. It is non-singular and its spatial gradients are well-defined \cite{JV2019}.

The scale-split magnetic energy density, $\psi({\bf{x}}, t)$ is defined \cite{JV2019} as
\begin{equation}\label{scale-splitB}
\psi({\bf{x}}, t)={1\over 2} \;{\bf{B}}_l({\bf{x}}, t){\bf{.B}}_L({\bf{x}}, t).
\end{equation}
which is divided into two scalar fields as $\psi({\bf x}, t)=\phi({\bf x}, t)\chi({\bf x}, t)$ such that
 
 \begin{equation}\label{phichi1}
\phi_{l,L} ({\bf{x}}, t)=\begin{cases}
\hat{\bf{B}}_l({\bf{x}}, t).\hat{{\bf{B}}}_L({\bf{x}}, t) \;\;\;\;\;\;B_L\neq 0\;\&\;B_l\neq0,\\
0\;\;\;\;\;\;\;\;\;\;\;\;\;\;\;\;\;\;\;\;\;\;\;\;\;\;\;\;\;\;\;\;\;otherwise,
\end{cases}
\end{equation}
which is called magnetic topology field and

\begin{equation}\label{phichi2}
\chi({\bf{x}}, t)={1\over 2} B_l ({\bf{x}}, t) B_L({\bf{x}}, t),
\end{equation}
which is called magnetic energy field. The quantity $\hat{\bf{B}}_l({\bf{x}}, t).\hat{{\bf{B}}}_L({\bf{x}}, t)$ is in fact the cosine of the angle between two coarse-grained components ${\bf B}_l$ and ${\bf B}_L$, hence it is a local measure of the field's stochasticity level. In order to develop a statistical measure, we can take the volume average of this quantity in a volume $V$ at time $t$ which defines magnetic stochasticity level $S_p(t)$ given by

\begin{equation}\label{formulae}
S_p(t)={1\over 2} ||\phi({\bf x}, t)-1 ||_p,
\end{equation}
where we have used the ${\cal L}_p$ norms for averaging\footnote{The ${\cal L}_p$ norm of ${\bf f}: {\cal{R}}^m\rightarrow {\cal{R}}^m$ is the mapping ${\bf f}\rightarrow ||{\bf{f}}||_p=[\int_V |{\bf f(x)} |^p (d^m x/V)]^{1/p}$. In this paper, we will take $p=2$ for simplicity, $||f||_2=f_{rms}$, which is the root-mean-square (rms) value of $\bf f$.  }. The cross energy is defined using the energy field $\chi({\bf x}, t)$ as

\begin{equation}\label{formulae2}
E_p(t)= ||\chi({\bf x}, t) ||_p,
\end{equation}

With $p=2$, the second order magnetic stochasticity level $S_2$, magnetic topological deformation $T_2=\partial_t S_2(t)$, magnetic cross energy density $E_2(t)$, and magnetic field dissipation $D_2=\partial_t E_2(t)$ are given by

\begin{equation}\label{formulae}
S_2(t)={1\over 2} (\phi-1  )_{rms},
\end{equation}

\begin{equation}
T_2(t)=  {1\over 4 S_2(t)} \int_V \;(\phi-1){\partial \phi\over \partial t}\; {d^3x\over V},
\end{equation}

\begin{equation}
E_2(t)=\chi_{rms},
\end{equation}

and
\begin{equation}
D_2(t)={1\over  E_2(t)}\int_V \chi \partial_t \chi{d^3x\over V}.
\end{equation}

\section{Diffusion in Turbulence}\label{s2}
In a resistive fluid, magnetic field lines will diffuse away as a result of a non-zero magnetic diffusivity $\eta$ (which is proportional to electrical resistivity). This phenomenon is similar to the diffusion of a passive scalar such as dye in a fluid like water. In Taylor diffusion (also called normal diffusion; the diffusion scheme present also in Brownian motion), the average (rms) distance of a particle from a fixed point, $\delta(t)$, increases with time $t$ as 
\begin{equation}\label{NormalDiff}
\delta^2(t)= D_T t,
\end{equation}
where $D_T$ is the (constant) diffusion coefficient. Note that no matter the medium is turbulent or not, this diffusion scheme will apply but with different diffusion coefficients. Turbulence will in general increase the diffusion coefficient $D_T$ making the diffusion process more efficient but the nature of this normal diffusion will remain linear in time (see below) at scales much larger than the turbulent inertial range. Because by definition $\delta^2\sim t^\gamma$ is sub-linear if $\gamma<1$, linear if $\gamma=1$ and super linear if $\gamma>1$, hence normal diffusion is a linear diffusion.

\begin{figure}
 \begin{centering}
\includegraphics[scale=.4]{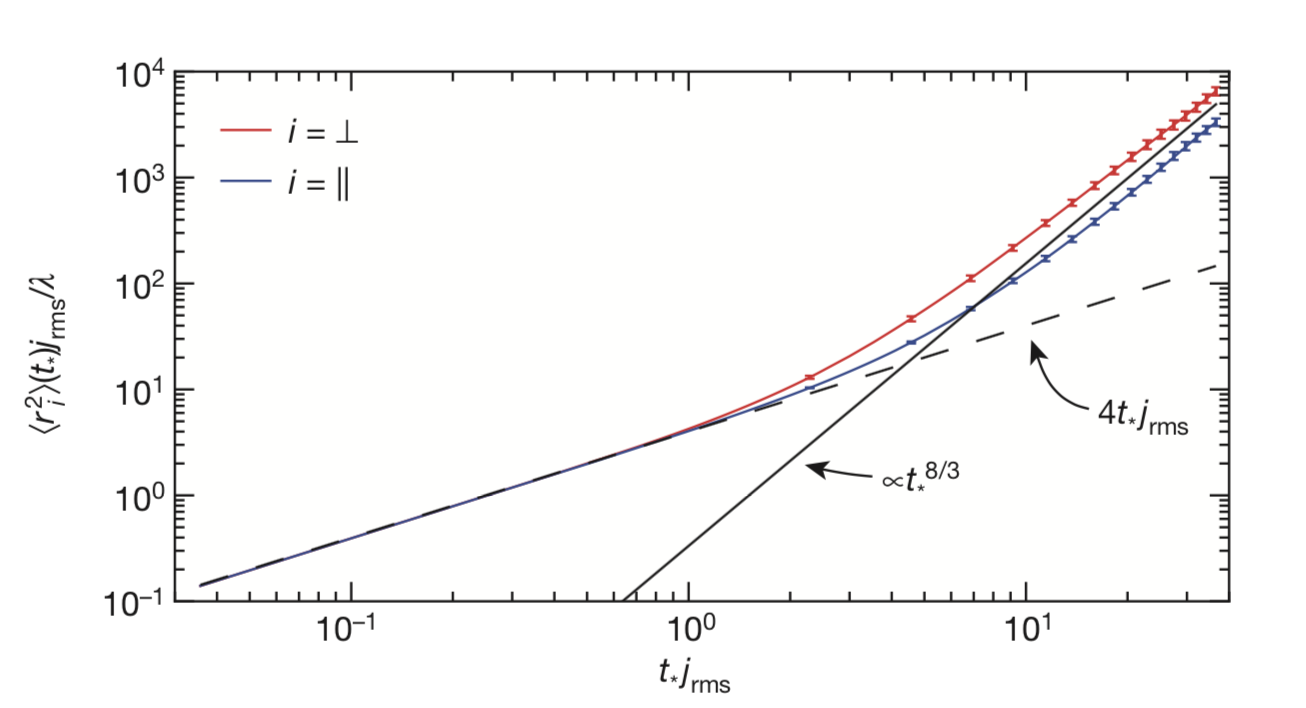}
\caption {\footnotesize {Evidence of super-linear magnetic dispersion in turbulence. Top: Mean squared dispersion of
field lines backwards in time with the variable $t_*=t_f-t$, in directions both
parallel (red) and perpendicular (blue) to the local magnetic field. Times are
normalized by the inverse r.m.s. current, $1/j_{rms}$, and distances are normalized by
the resistive length, $(\lambda/j_{rms})^{1/2}$. Error bars, s.e.m. The dashed line shows the conventional diffusive estimate, $\langle r^2(t_*)\rangle=4\lambda t_*$, and the solid line is $\propto t_*^{8/3}$. For details see \cite{Eyinketal.2013}.}}\label{Richardson}
\end{centering}
\end{figure}

In the presence of turbulence, although the normal diffusion scheme is still valid at scales much larger than the large eddies in the inertial range but it cannot be applied in the inertial range of turbulence. In the inertial range, the average (rms) separation of two diffusing particles grows super-linearly with time. This corresponds to 2-particle Richardson diffusion;
\begin{equation}\label{RichDiff}
\Delta^2(t)= D_R t^3,
\end{equation}
with diffusion coefficient $D_R$. This result can be obtained in several ways discussed in the Appendix (see also \cite{Eyinketal.2013}; \cite{JafariandVishniac2018}). The power of $3$ indicates, of course, a super-linear diffusion. It is important to emphasize that the Richardson diffusion is a 2-particle diffusion (i.e., it is concerned with the average separation of two particles undergoing diffusion in turbulence) while the normal (Taylor) diffusion is a one-particle diffusion scheme (i.e., it is concerned with the average distance of a diffusing particle from a fixed point). It turns out, as it might be expected, that magnetic field lines undergo Richardson diffusion in the turbulence inertial range \cite{Eyinketal.2013}; see Fig.(\ref{Richardson}).

Spectral analysis (Fourier decomposition) is often used to study turbulent magnetic fields in which one speaks of parallel $\lambda_\parallel=k_\parallel^{-1}$ and perpendicular $\lambda_\perp=k_\perp^{-1}$  wave-lengths and wave-numbers ($k_\parallel$ and $k_\perp$) with respect to the local magnetic field. In such an approach, general equations (\ref{NormalDiff}) and (\ref{RichDiff}) are translated into the following relationship:
\begin{equation}\label{generalL}
\lambda_\perp^2\approx \alpha \lambda_\parallel^\beta,
\end{equation}
with fixed $\alpha$ (for a given turbulence inertial range) as the diffusion coefficient; see Appendix. Note that $\beta=3$ corresponds to the super-linear Richardson diffusion and $\beta=1$ to normal, sub-linear dissipative diffusion. How can we relate magnetic diffusion to the level of randomness in magnetic field?

\begin{figure}[h]
 \begin{centering}
\includegraphics[scale=.23]{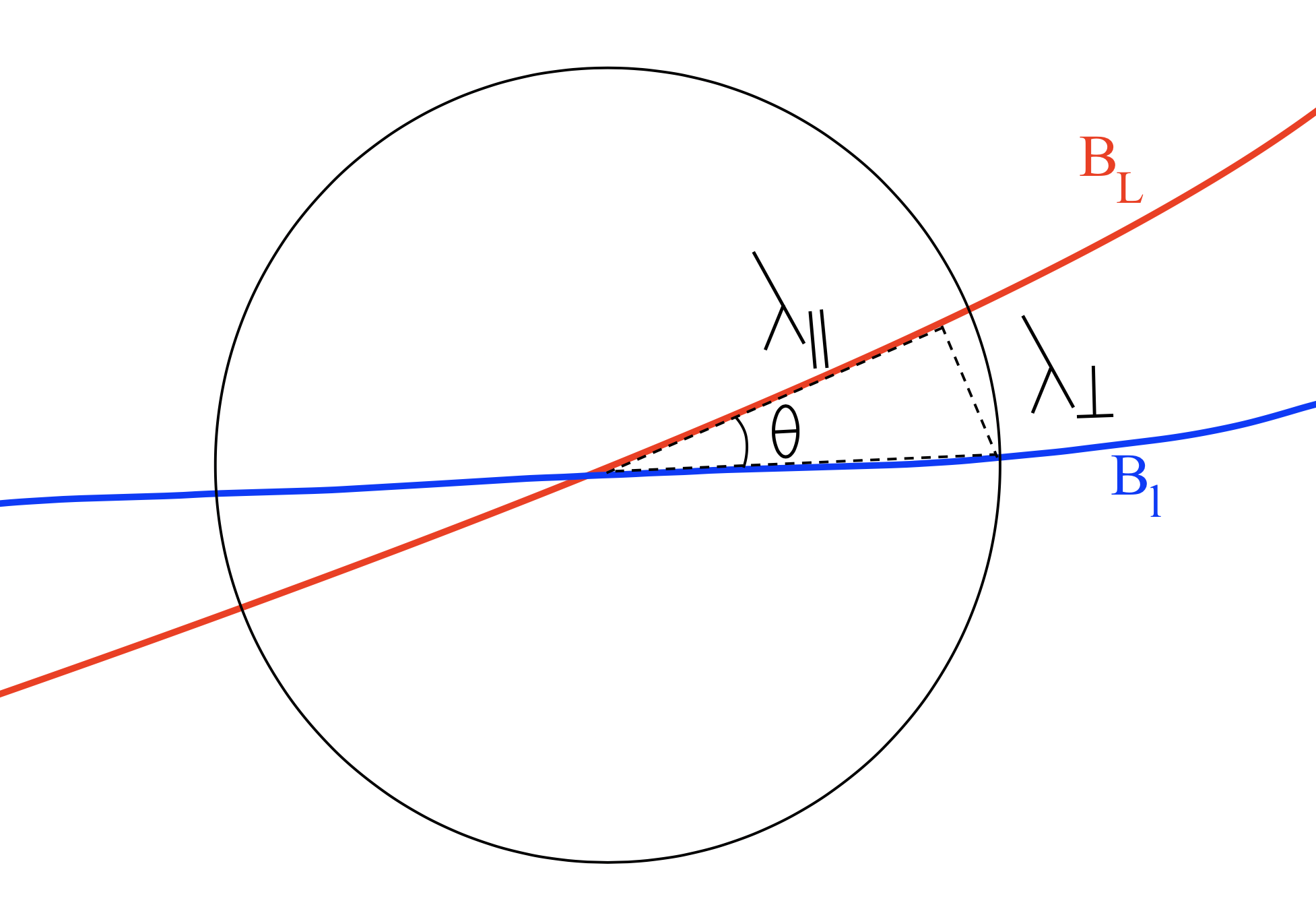}
\caption {\footnotesize {Parallel, $\lambda_\parallel$, and perpendicular, $\lambda_\perp$, wavelengths with respect to the large scale magnetic field ${\bf B}_L$ (coarse-grained $\bf B$ on scale $L$), in a region of scale $l<L$ with local field ${\bf B}_l$ (coarse-grained $\bf B$ on scale $l<L$). The angle $\theta$ is a stochastic variable because of the randomness inherent in $\bf B$, hence it can be used to quantify magnetic stochasticity level. Its local relationship with $\lambda_\parallel$ and $\lambda_\perp$, on the other hand, provides a means to relate the stochasticity level $S_2(t)$, defined by eq.(\ref{formulae}), to magnetic diffusion, eq.(\ref{generalL}). This relationship is quantified by eq.(\ref{main2}).}}\label{Diff}
\end{centering}
\end{figure} 

Consider the coarse-grained magnetic field ${\bf B}_l$ in a region of length scale $l$. Denote by $\lambda_\perp$ and $\lambda_\parallel$, respectively, the perpendicular and parallel components of ${\bf B}_l$ with respect to ${\bf B}_L$; see Fig.(\ref{Diff}). The assumption is that we are in the inertial range of turbulence and $L$ is few times larger than $l$. From eq.(\ref{generalL}) one can write
\begin{equation}\notag
\lambda_\parallel=l \phi,\;\;\&\;\;\lambda_\perp=l (1-\phi^2)^{1/2},
\end{equation}

where $\phi({\bf x}, t)=\cos\theta=\hat{\bf B}_l.\hat{\bf B}_L$. We have
\begin{equation}\label{111}
l^2(1-\phi^2)=\alpha l^\beta\phi^\beta.
\end{equation}

One can relate the stochasticity level, given by eq.(\ref{formulae}), to the magnetic diffusion, which is related to eq.(\ref{111}), by writing the latter expression as
\begin{equation}\notag
{1\over 2}(1-\phi)={\alpha\over 2} l^{\beta-2}{\phi^\beta\over 1+\phi}, 
\end{equation}
which upon taking ${\cal L}_p$-norm and using eq.(\ref{formulae}) gives us

\begin{equation}\label{main1}
S_p(t)={\alpha\over 2} l^{\beta-2}\parallel{\phi^\beta({\bf x}, t)\over 1+\phi({\bf x}, t)}\parallel_p.
\end{equation}

Taking $p=2$, for simplicity, we find 
\begin{equation}\label{main2}
S_2(t)={\alpha\over 2} l^{\beta-2} \Big({\phi^\beta({\bf x}, t)\over 1+\phi({\bf x}, t)}  \Big)_{rms}.
\end{equation}

\begin{table}[ht]
\caption{Numerical values of $\langle f(t)\rangle_T$ for different scales $l$ and $L$, which is assumed to be few times larger than $l$, for $\beta=1, 3, 5, 7$ in a randomly selected sub-volume of size $194\times 42\times 33$ in grid units; see eq.(\ref{Avef}). The mean, standard deviation (STD) and relative standard deviation of $\langle f(t)\rangle_T$ are calculated for different scales $l$ and $L$. These data, and similar ones for other randomly selected sub-volumes, indicate that $\beta=3$ gives the smallest relative standard deviation for $\langle f(t)\rangle_T$. Physically, this means that only for super-diffusion with $\beta\sim 3$ an almost constant diffusion coefficient can be obtained as $\alpha=2\langle f(t)\rangle_T$.}
\centering
\scalebox{1.07}{
\begin{tabular}{c c c c c c c c c c}
\hline\hline
& $l,\;L$ &\;\;& $\beta=1$ &\;\;& $\beta=3$ &\;\;& $\beta=5$ &\;\;& $\beta=7$ \\ [0.5ex] 
\hline
&$3, 7$&\;\;\;\;&0.2774&\;\;&0.0317&\;\;&0.0036&\;\;&0.0004\\
&$3, 9$&\;\;\;\;&0.2128 &\;\;&0.0244&\;\;&0.0028&\;\;&0.0003 \\
& $5, 9$ &\;\;\;\;& 0.5834 &\;\;&0.0241&\;\;&0.0009&\;\;&0.0000 \\ 
& $3, 11$ &\;\;\;\;& 0.1590 &\;\;&0.0180&\;\;&0.0020&\;\;&0.0002 \\ 
& $5, 11$&\;\;\;\;& 0.5087 &\;\;&0.0212&\;\;&0.0008&\;\;&0.0000 \\ 
&$7, 11$ &\;\;\;\;& 0.8985 &\;\;& 0.0189&\;\;&0.0004&\;\;&0.0000 \\ 
 &Mean &\;\;\;\;&0.4400&\;\;&0.0230&\;\;&0.0018 &\;\;& 0.0002  \\ 
 & STD & \;\;\;\;&0.2802 &\;\;&0.0050&\;\;&0.0012&\;\;&0.0002 \\ 
 & Relative STD &\;\; \;\;&0.6368&\;\;&0.2174&\;\;&0.6667&\;\;&1.000 \\ [1ex]
\hline
\end{tabular}
}
\label{table1}
\end{table}

First, we note that the scale dependence decreases with increasing the scale since we are coarse-graining using a rapidly decaying kernel $G$. Hence with $L> l$, we expect a weaker dependence on $L$ in eq.(\ref{main2}), which is based on the assumption that $L$ is few time larger than $l$. The relationship given by eq.(\ref{main2}) should hold for the right choice of $\beta$ for magnetic diffusion (i.e., $\beta=1$ for resistive diffusion and $\beta=3$ for Richardson diffusion). With such a choice, the diffusion coefficient $\alpha$ can be obtained from this expression by time-averaging. If the diffusion scheme is super-linear Richardson diffusion in the inertial range, as discussed above, the evaluation of the above expression should lead to $\beta=3$ with a fixed diffusion coefficient $\alpha$.


\begin{figure}
 \begin{centering}
\includegraphics[scale=.415]{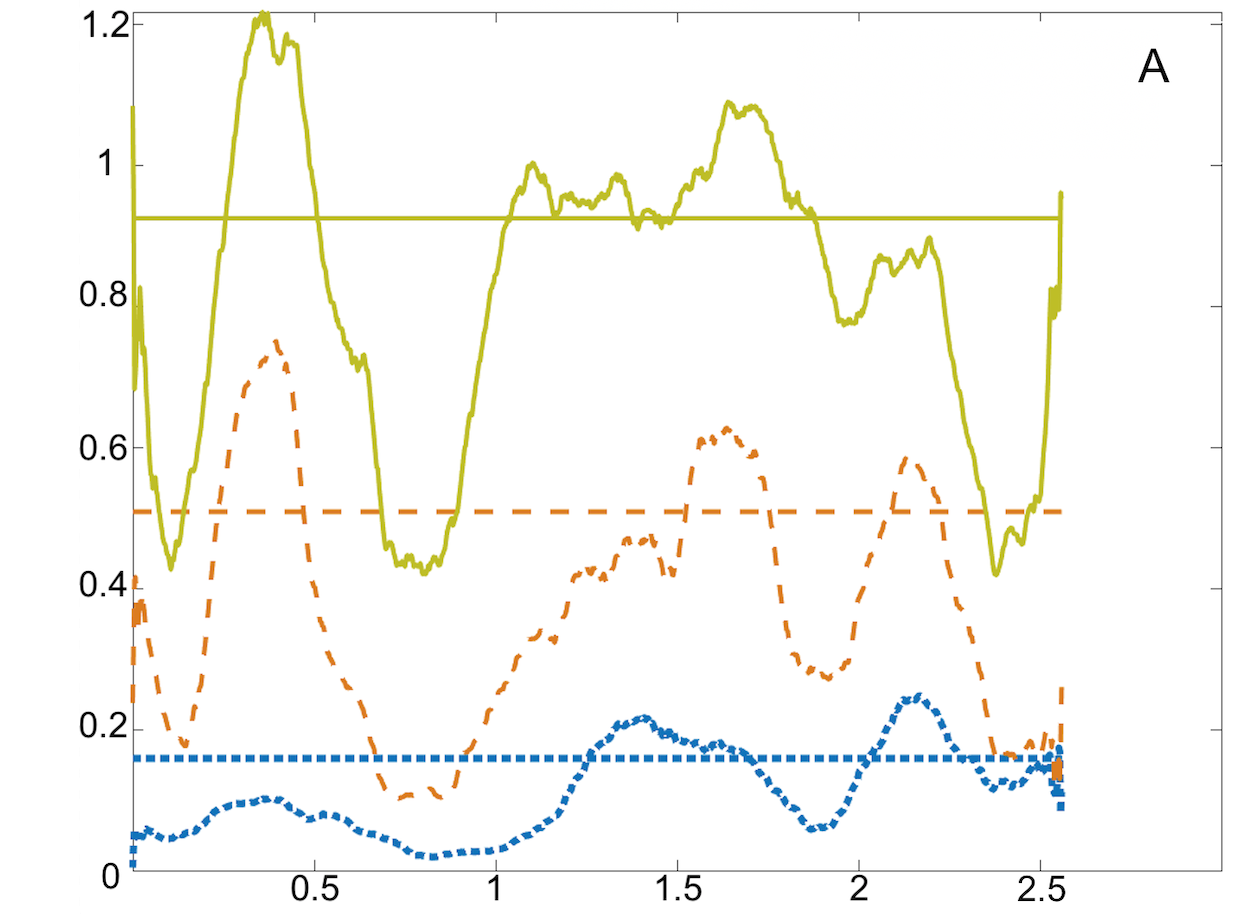}
 \includegraphics[scale=.4]{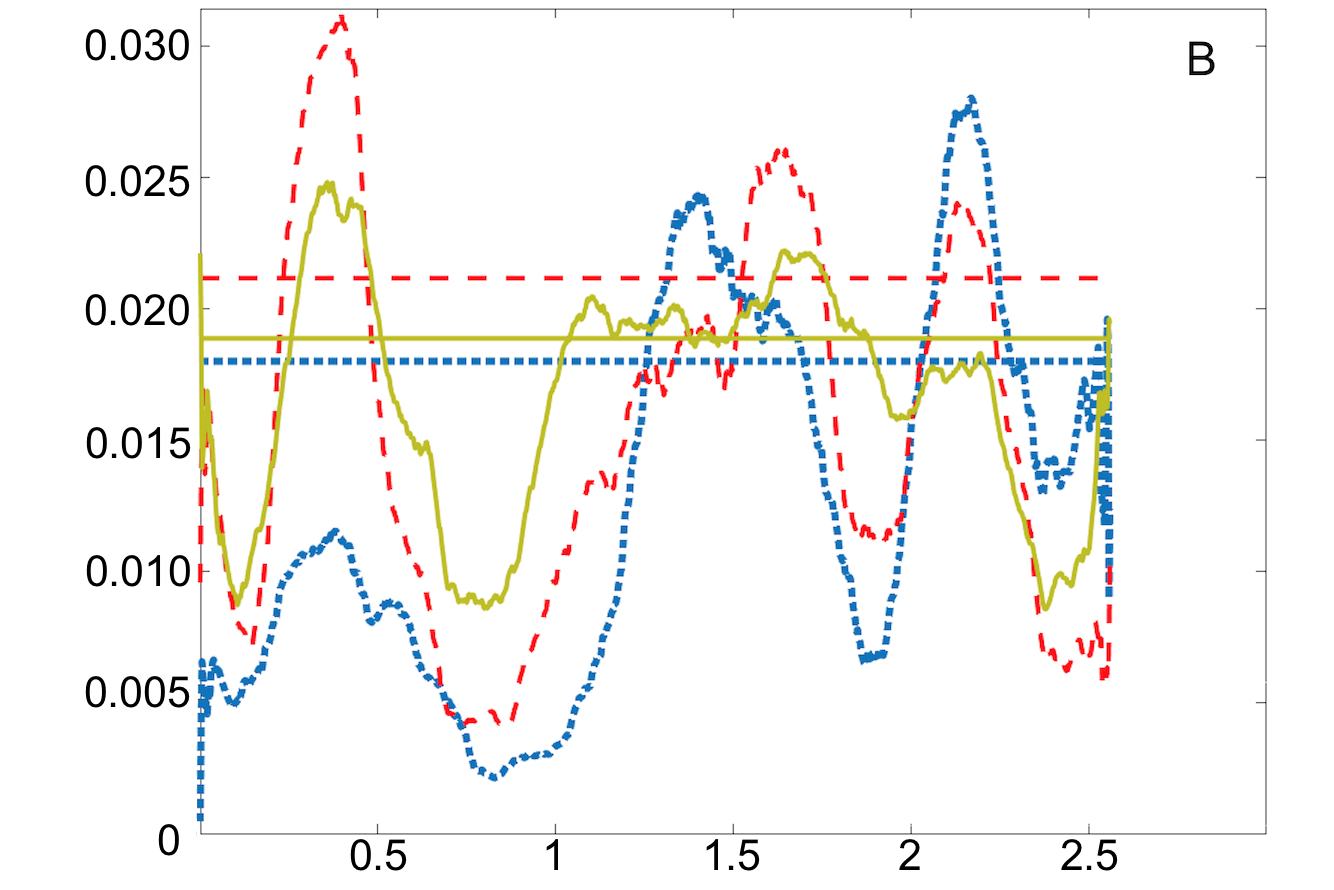}
\caption {\footnotesize {Plots of $f(t)$ (curves) defined by eq.(\ref{N1}) with respect to time and its time-average $\langle f(t)\rangle_T$ (horizontal lines) for $l=3, L=11$ (dotted, cyan), $l=5, L=11$ (dashed, red), $l=7, L=11$ (solid, yellow). For the correct value of $\beta$, the time average of this function, $\langle f(t)\rangle_T$, should be almost independent of scale and approximately equal to the half of the diffusion coefficient $\alpha$, defined by eq.(\ref{generalL}). For different scales, the standard deviation and relative standard deviation corresponding to $\beta=1$ (A) are much larger that their counterparts for $\beta=3$ (B). The numerical values of $\alpha$, in the same sub-volume, are shown in Table.(\ref{table1}). This result holds in different sub-volumes of the simulation box.}}\label{DiffusionPlots1}
\end{centering}
\end{figure}

\begin{figure}
 \begin{centering}
\includegraphics[scale=.355]{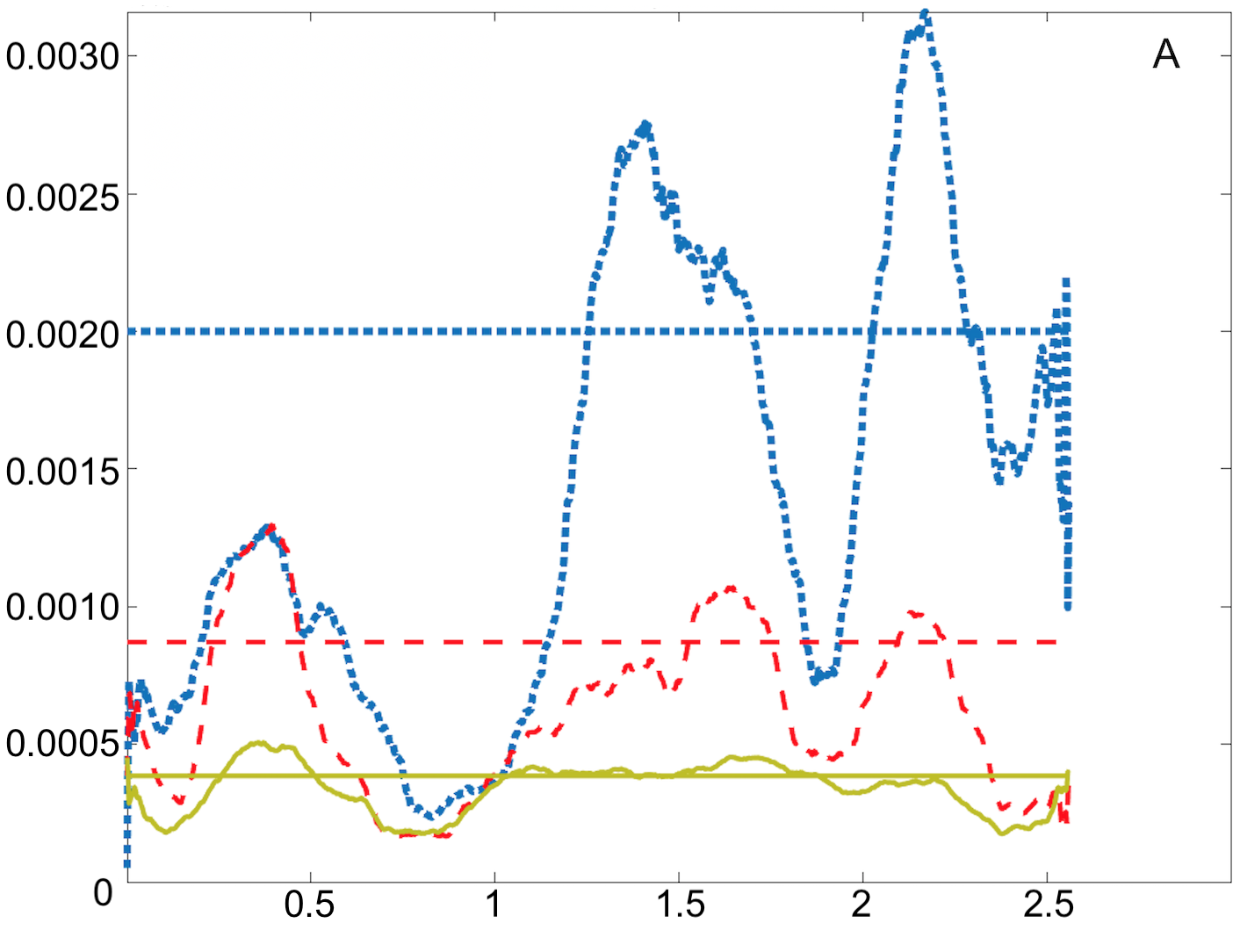}
\includegraphics[scale=.33]{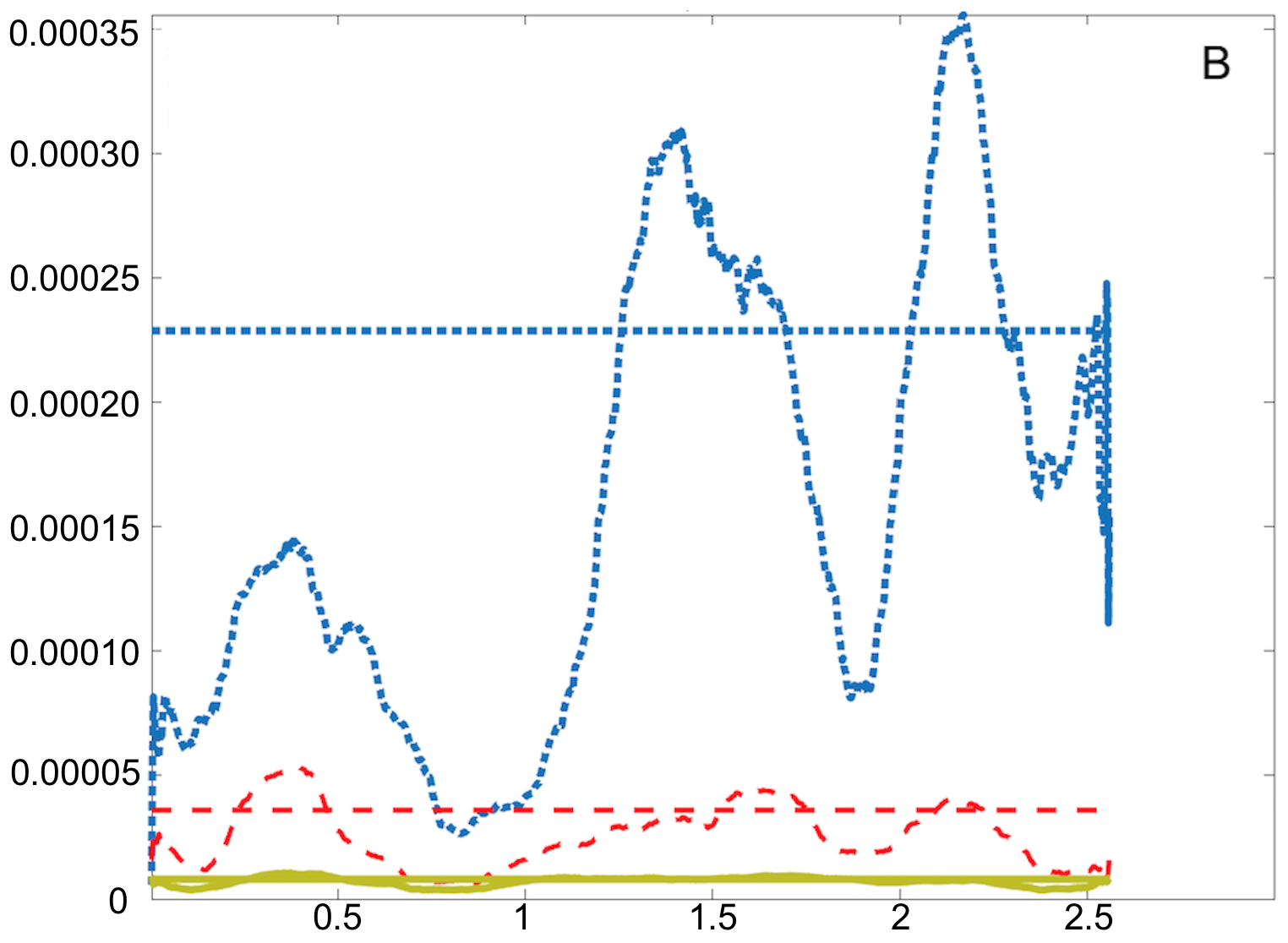}
\caption {\footnotesize {Same as Fig.(\ref{DiffusionPlots1}) but for $\beta=5$ (A) and $\beta=7$ (B); see also Table.(\ref{table1}). The value of $f(t)$ becomes smaller by increasing $\beta$, however, its relative standard deviation ramps up indicating that $f(t)$ is not a constant for $\beta> 3$ and thus no constant diffusion coefficient can be defined. For $\beta\neq 3$, in general, a slight change in scale leads to a large relative change in diffusion coefficient.}}\label{DiffusionPlots2}
\end{centering}
\end{figure} 
In order to examine eq.(\ref{main2}) numerically, we use a homogeneous, incompressible MHD numerical simulation archived in an online, web-accessible database (\citep*{JHTDB};\citep*{JHTB1};\citep*{JHTB2}). This is a direct numerical simulation (DNS), using $1024^3$ nodes, which solves incompressible MHD equations using pseudo-spectral method. The simulation time is $2.56\;s$ and $1024$ time-steps are available (the frames are stored at every $10$ time-steps of the DNS). Energy is injected using a Taylor-Green flow stirring force. Let us define 
\begin{equation}\label{N1}
f(t):={1\over l^{\beta-2}}{S_2(t)\over \Big({\phi^\beta({\bf x}, t)\over 1+\phi({\bf x}, t)}\Big)_{rms}},
\end{equation}
and evaluate the (rms) time average of this function, $\langle f(t)\rangle_T$ over the time interval $T$ (simulation time). For the right choice of $\beta$ for magnetic diffusion in turbulence, if eq.(\ref{N1}) holds, $\langle f(t)\rangle_T$ should be a constant and independent of $l$:

\begin{equation}\label{Avef}
\langle f(t)\rangle_T={\alpha\over 2}=const.
\end{equation}

In Fig.(\ref{DiffusionPlots1}), we have plotted $f(t)$ as a function of time, for different values of $l$ and $L$ for both $\beta=1$ and $\beta=3$ in a randomly selected sub-volume of the simulation box with size $194\times 42\times 33$ in grid units equivalent to $1.2\times 0.26\times 0.20$ in physical units with coordinates $[400,733,300]-[596,775,333]$ in the simulation box. Repeating this computation in several other randomly selected sub-volumes of the simulation box leads to similar results. We have also evaluated other values of $\beta$ in the same sub-volume; e.g., see Fig.(\ref{DiffusionPlots2}) for $\beta=5, 7$. The numerical values of $\langle f(t)\rangle_T$ for different scales $l$ and $L$, along with the mean value, standard deviation and relative standard deviation, are presented in Table.(\ref{table1}). These results indicate that $\beta=3$ (Richardson diffusion) leads to an almost constant (with respect to scales $l$ and $L>l$) diffusion coefficient $D_R:=\alpha$ while the relative standard deviation of the time-averaged quantity $\langle f(t)\rangle_T$ becomes increasingly larger as $\beta$ increases and no constant diffusion coefficient, with respect to scale, can be defined in these cases.

\section{Summary and Conclusions}\label{s3}

In this paper, we have advanced physical arguments to relate magnetic stochasticity level $S_p(t)$ to magnetic dispersion in MHD turbulence. We have also tested this theoretical prediction using an incompressible, homogeneous MHD numerical simulation stored online (\citep*{JHTDB};\citep*{JHTB1};\citep*{JHTB2}). Our results agree with the super-linear, Richardson diffusion scheme for turbulent magnetic fields.

Stochasticity level of turbulent magnetic fields is quantified by volume-averaging the scalar field $\phi({\bf x}, t)=\hat{\bf B}_l.\hat{\bf B}_L$ where $\hat{\bf B}_l={\bf B}_l/|{\bf B}_l|$ and ${\bf B}_l$ is the coarse-grained magnetic field at scale $l$; ${\bf B}_l=l^{-3} \int_V G({\bf r}/l) {\bf B}({\bf x+r}, t)d^3r$ with a rapidly decaying kernel $G({\bf r})=G(r)$. Likewise $\hat{\bf B}_L$ can be computed for a larger scale $L>l$. More specifically, $p$-order magnetic stochasticity is defined as the ${\cal L}_p$-norm (volume-average) $||\phi-1||_p/2$. Hence for $p=2$, the second order magnetic stochasticity level is given by the rms-average $S_2(t)={1\over 2}(\phi-1)_{rms}$ \cite{JV2019}. 

We have shown, using simple scaling laws of MHD turbulence, that stochasticity level $S_p(t)$ is related to the diffusion power $\beta$ and diffusion coefficient $\alpha$ in $\lambda_\perp^2\approx \alpha \lambda_\parallel^\beta$ by eq.(\ref{main1});
\begin{equation}\notag
S_p(t)={\alpha\over 2} l^{\beta-2}\parallel{\phi^\beta({\bf x}, t)\over 1+\phi({\bf x}, t)}\parallel_p.
\end{equation}

We have used the second order stochasticity level $S_2(t)$ to numerically check the Richardson value $\beta=3$ against normal diffusion associated with $\beta=1$. In our statistical analyses of several sub-volumes of the simulation box, super-linear diffusion with $\beta=3$ leads to the smallest standard deviation, which is at least an order of magnitude smaller than that corresponding to $\beta=1$. It should be emphasized that we have related magnetic stochasticity to its diffusion scheme in turbulence, whose super-linear nature can be inferred using any MHD turbulence model such as Goldreich-Sridhar model \cite{GoldreichandSridhar1995} or Kolmogorov scaling laws \cite{Kolmogorov1941}, discussed in the Appendix. Our arguments in this paper are thus quite generally applicable to magnetic fields in turbulence inertial range and are independent of any MHD turbulence model. 

\appendix
\section{Richardson Diffusion and MHD Turbulence}

In this Appendix, we present well-known theoretical evidence in support of the super-linear nature of magnetic diffusion in the presence of turbulence invoking different methodologies. The super-linear nature of magnetic diffusion does not depend on any MHD turbulence model, rather the implication is that any successful MHD turbulence model would agree with Richardson 2-particle diffusion scheme. Super-linear magnetic diffusion in turbulence inertial range is a model-independent, universal feature of turbulent magnetic field. Our arguments in this paper relate this phenomenon to the stochasticity level of magnetic fields. Analytical and numerical studies of magnetic diffusion in turbulence inertial range can be found in e.g., \cite{Eyink2011}; \cite{Eyinketal.2013}; \cite{Lalescuetal.2015}; \cite{JafariandVishniac2018} and \cite{Eyink2019} and references therein.

To estimate the 2-particle separation in a turbulent flow (for a detailed discussion see \cite{Eyink2019}), we write the distance between two arbitrary particles, initially separated by $\Delta(t=0)\equiv \Delta_0=\Delta {\boldsymbol{\alpha}}_0$, at time $t$ as
\begin{equation}\label{R1}
\Delta(t)=|{\bf X}({\boldsymbol{\alpha}}', t)-{\bf X}({\boldsymbol{\alpha}}, t)   |,
\end{equation}
where ${\boldsymbol{\alpha}}'={\boldsymbol{\alpha}}+\Delta {\boldsymbol{\alpha}}_0$. Assuming a H{\"o}lder singular velocity field $\bf u$, we have
\begin{equation}\label{R2}
|{\bf u}({\bf x}', t)-{\bf u}({\bf x}, t)|\leq A|{\bf x}'-{\bf x}|^h,
\end{equation}
where $A$ is a constant and $h<1$ is the H{\"o}lder exponent. Taking the time derivative of eq.(\ref{R1}), using the triangular inequality and eq.(\ref{R2}), we find
\begin{equation}
{d\Delta (t)\over dt}\leq A |  {\bf X}({\boldsymbol{\alpha}}', t)-{\bf X}({\boldsymbol{\alpha}}, t)  |^h=A[\Delta(t)]^h,
\end{equation}
with the solution

\begin{equation}
\Delta(t)\leq\Big[ \Delta_0^{1-h}+A(1-h)(t-t_0)  \Big]^{1\over 1-h}.
\end{equation}

There is a remarkable difference in the above expression for two different choices $0<h<1$ and $h\rightarrow 1$. It is simple calculus to show that the latter case leads to 
\begin{equation}
\Delta(t)\leq \Delta_0 \exp{ [A(t-t_0)  ]},\;\;h\rightarrow 1.
\end{equation}
This result implies that for $t\rightarrow \infty$, we have $\Delta(t)\simeq \Delta_0e^{A(t-t_0)}$. Therefore we arrive at the important result that in this case the initial conditions are never forgotten! However, the choice $0<h<1$ leads to 
\begin{equation}
\Delta(t)\simeq \Big[ A(1-h)(t-t_0) \Big]^{1\over 1-h},\;\;t\rightarrow\infty.
\end{equation}
This implies that the information about initial conditions, i.e., the initial separation of particles, is lost for long times.
Using the Kolmogorov's (\cite{Kolmogorov1941}) theory of turbulence in the inertial range, we know that $h=1/3$. Therefore, the above expression yields 

\begin{equation}\Delta^2(t)\propto (t-t_0)^{3}.
\end{equation}
The power of $3$ indicates, of course, a super-linear diffusion. At sufficiently large times $t\gg t_0$, we recover eq.(\ref{NormalDiff}): $\Delta^2(t)\sim t^3$.

Historically, however, Richardson took a different approach to get this result, an understanding of which is both instructive and also important for many other problems (see  also \cite{Eyink2015}). Richardson's probability density for particle separation vector ${\bf{l}}={\bf{x}}_1-{\bf{x}}_2$, with a scale-dependent diffusion  coefficient $K(l)\sim K_0 l^{4/3}$, satisfies $\partial_t P({\bf{l}},t)=\nabla_{l_i}\Big( K(l) \nabla_{l_i} P({\bf{l}},t)\Big)$ with a similarity solution \cite{Eyink2011},

\begin{equation}\label{Richardson1}
P({\bf{l}},t)={A\over (K_0 t)^{9/2} }\exp{\Big( -{9l^{2/3}\over 4 K_0 t} \Big) }.
\end{equation}
Using this probability density to average $l^2$, one finds $\langle l^2(t)\rangle=(1144/81) K_0^3 \; t^3$. This is intimately related to Kolmogorov's relation 
\begin{equation}\label{t3}
l^2(t)\sim (g_0 \epsilon) t^3,
\end{equation} 
which is a solution to the initial value problem $dl(t)/dt=\delta u(l)=(3/2)(g_0\epsilon l)^{1/3}$, $l(0)=l_0$ for sufficiently long times $t\gg t_0$. Here $g_0$ is Richardson-Obukhov constant and $\epsilon$ the mean energy dissipation rate.

In the following we also present a brief discussion of MHD turbulence scaling laws related to magnetic diffusion in the turbulence inertial range. We shall see that such considerations generally agree with the reulst given by eq.(\ref{t3}).

In the Kolmogorov's hydrodynamic turbulence \cite{Kolmogorov1941}, the statistics of turbulent motions is determined by the energy transfer rate $\epsilon=v_y^3/y$ and length scale $y$ at the inertial range, where the energy is supposedly transferred with no dissipation, and the energy transfer rate and viscosity $\nu$ at the dissipative range, where the energy is dissipated by viscosity. Dimensional analysis leads to the turbulent velocity $v_y\sim \epsilon^{1/3}y^{1/3}$ and the turbulent eddy time scale $\tau_y\sim \epsilon^{-1/3}y^{2/3}$ in the inertial range. 
This scaling leads to the famous velocity spectrum $v_k\sim k^{-1/3}$ and energy power spectrum 

\begin{equation}\label{Kscaling}
E_{Kol}(k)\sim k^{-5/3}. 
\end{equation}
At the dissipative range, a similar dimensional analysis shows that $v_d\sim \epsilon^{1/4}\nu^{1/4}$, $\tau_d\sim \epsilon^{-1/2}\nu^{1/2}$ and $y_d\sim \nu^{3/4}\epsilon^{-1/4}\equiv y(\nu/y v_y)^{3/4}$. 

Kolmogorov scaling, for hydrodynamic turbulence, cannot be applied directly to MHD turbulence because of the complications that magnetic field introduces. Iroshinkov \cite{Iroshnikov1963}, and Kraichnan \cite{Kraichnan1965} independently developed a model for incompressible MHD turbulence. Since two-wave interactions behave as elastic collisions, this picture relies on the interactions of triads of waves \cite{DiamondandCraddock1990}. The Iroshinkov-Kraichnan (IK) model assumes that (a) only oppositely directed waves interact, (b) turbulence is isotropic and (c) energy cascades from long to short wavelengths, and finally (d) dominant interactions involve three wave-couplings (triads). The assumption that interactions are local, i.e. only fluctuations of comparable sizes interact, then propagating fluctuations behave as Alfv\'en wave packets of parallel extent $l_{\parallel}$ and perpendicular extent $l_{\perp}$. Suppose $\delta u_{l} \sim \delta B_{l}$, where $\delta u_{l}$ and $\delta B_{l}$ are, respectively, the fluctuations in the velocity and magnetic fields. Two counter-propagating wave packets would require an Alfv\'en time, $\tau_{A} \sim l_{\parallel}/V_A$, to pass through each other. During this time, the amplitude of each wave packet will suffer a change $\Delta \delta u_{l}$,

\begin{equation}\label{61-3}
{\Delta \delta u_l\over \tau_A} \sim {\delta u_l\over \tau_l},
\end{equation}

where $\tau_{l} \sim l/ \delta u_{l}$ is the eddy turnover time. The assumption of the local, weak interactions means
\begin{equation}\label{61-2}
\Delta \delta u_{l} \ll \delta u_{l} \Leftrightarrow \tau_{A} \ll \tau_{l}.
\end{equation}
The cascade time, $\tau_{nl}$ is defined as the time that it takes to change $\delta u_{l}$ by an amount comparable to itself assuming that the changes accumulate in a random walk manner\footnote{The average total length in a random walk with average step size $\overline l$ vanishes $\langle\Sigma_i l_i\rangle=0$, but  $\langle(\Sigma_i l_i)^2\rangle=\langle\Sigma_i l_i^2\rangle=N \overline l^2$. The total number of steps $N$ is the total time $t_{tot}$ divided by the average time of one step $\overline t$, that is  $N=t_{tot}/\overline t$. Thus  $\langle(\Sigma_i l_i)^2\rangle^{1/2}=(t_{tot}/\overline t)^{1/2} \overline l$. For the velocity fluctuations, eq.(\ref{61-3}) gives $\Delta \delta u_l\sim \delta u_l (\tau_A/\tau_l)$. To add up $\Delta \delta u_l$'s during total time $\tau_{nl}$, that is $\sum \limits_{t=0}^{\tau_{nl}}\Delta \delta u_l=(\tau_A/\tau_l) \sum \limits_{t=0}^{\tau_{nl}}\delta u_l $, one replaces the step size with $\delta u_l$, average step time with $\tau_A$. This leads to the first part of eq.(\ref{61-1}).};
\begin{equation}\label{61-1}
\sum \limits_{t=0}^{\tau_{nl}} \; \Delta \delta u_{l} \sim \delta u_{l} \frac{\tau_{A}}{\tau_{l}} \left({\frac{\tau_{nl}}{\tau_{A}}}\right)^{1/2} \sim \delta u_{l}.
\end{equation}
Thus, we obtain the energy transfer rate
\begin{equation}\label{61-0}
\tau_{nl} \sim \frac{l^2 V_{A}}{l_{\parallel} \delta u_{l}^2}.
\end{equation}

The isotropy assumption means that the power is distributed isotropically in wave space. Assuming $l_{\parallel} \sim l$, we obtain
  \begin{equation}\label{61}
\tau_{IK} \simeq \frac{l V_A}{v_k^2},
\end{equation}
where we have simply used $v_{k}$, to refer to the rms velocity on the eddy scale $k^{-1}$, instead of $\delta u_{l}$. When $v_k < V_A$, we have $\tau_{nl} > k v_k$, which implies the reduction of the energy cascade to higher wavenumbers by the magnetic field. Constancy of the energy transfer rate $v_k^2/\tau_{IK}=const.$ leads to $v_k\sim k^{-1/4}$. Therefore the Kraichnan-Iroshinkov energy power spectrum is given by 

\begin{equation}\label{IKScaling}
E_{IK}(k)\sim k^{-3/2}.
\end{equation}

IK theory was the most popular model accepted as MHD generalization of Kolmogorov's ideas for about 30 years. In 1970's, measurements showed strong anisotropies in the solar wind with $l_{\parallel} > l_{\perp}$. Goldreich and Sridhar (\cite{GoldreichandSridhar1995} henceforth GS95; \cite{GoldreichandSridhar1997}) suggested that the effect of residual three wave couplings is consistent with eq.(\ref{61-0}), for the basic nonlinear timescale, but an anisotropic spectrum should be considered in which the transfer of  power between modes moves energy toward larger $k_{\bot}$ with no effect on $k_{\parallel}$. Here, $k_{\bot}$ is the wavevector component perpendicular, and $k_{\parallel}$ is the wavevector component parallel, to the direction of magnetic field. Therefore, using eq.(\ref{61-0}), the basic nonlinear timescale can be written as
\begin{equation}\label{62}
\tau_{nl} \simeq \frac{k_{\parallel}V_A}{k_{\bot}^2 v_k^2},
\end{equation}
where $\omega_A=k_{\parallel}V_A$ is Alfv\'en wave frequency.

 The critical balance requires that $k_\|$ and $k_\perp$ are related as
\begin{equation}\label{critical-balance}
k_\| V_A\approx k_\perp v_k,
\end{equation}
where $V_A$ is the Alfv\'en speed and $v_k$. This is translated into the requirement that the field couples to a typical eddy at a rate approximately equal to the eddy turnover rate. The second assumption in the GS95 model is that the nonlinear energy transfer rate is $\sim k_\perp v_k$, similar to that of hydrodynamic turbulence \cite{Kolmogorov1941}. These assumptions together lead to a power spectrum which behaves like hydrodynamic turbulence, i.e. $v_k\propto k_\perp^{-1/3}$. Consequently, the energy power spectrum of GS95 is given by

\begin{equation}\label{GS95Scaling}
E_{GS}(k_\perp)\sim k_\perp^{-5/3}.
\end{equation}

Suppose that we inject energy into the medium with a parallel length scale $l$, with the corresponding perpendicular scale $l_{\bot}=l_\parallel M_A$ (where $M_A=V_T/V_A$ is the Alfv\'en Mach number), that creates an rms velocity $V_T$. The resulting inertial turbulent cascade satisfies critical balance at all smaller scales, therefore
\begin{equation}
\tau_{nl}^{-1}\simeq k_\| V_A\simeq k_{\perp}v_k,
\end{equation}
and the constant flow of energy through the sub-Alfv\'enic cascade is given by
\begin{equation}\label{Vrate}
\epsilon\simeq {v_l^4\over l_\parallel V_A}\simeq {V_T^2 \over l_\parallel/V_A}\simeq {v_k^2\over \tau_{nl}}\simeq k_\perp v_k^3,
\end{equation}
where $v_l=\sqrt{V_T V_A}$ is sometimes defined as the velocity for isotropic injection of energy which undergoes a weakly turbulent cascade and ends up with a strongly turbulent cascade (see e.g., \cite{LazarianandVishniac1999}; \cite{Jafarietal2018}). Putting all this together, we find the following relationship:

\begin{equation}\label{67}
k_{\parallel}\simeq l_\parallel ^{-1}\left({k_\perp l V_T\over V_A}\right)^{2/3},
\end{equation}
between parallel and perpendicular wave-numbers. Note that from here we also get $\tau_{nl}^{-1} \simeq k_\| V_A\simeq  {V_A\over l_\parallel }\left({k_\perp l V_T\over V_A}\right)^{2/3}$ and the rms velocity in the large scale eddies, $ v_k \simeq  V_T\left({k_\perp l V_T\over V_A}\right)^{-1/3}$.

Thus the perpendicular wave-number $k_\perp$ scales as $k_\perp\propto k_\parallel^{3/2}$ in terms of the parallel wave-number $k_\parallel$. This scaling is exactly similar to the Richardson scaling given by eq.(\ref{t3}). This result can be presented in terms of the wave-lengths parallel and perpendicular to the local magnetic field as $\lambda^2_\perp\sim \lambda_\parallel^3$.

\bibliographystyle{apsrev4-2}
\bibliography{SuperlineararXiv}

\end{document}